\documentstyle[psfig]{mn}

\title{Gamma Ray Bursts with peculiar temporal asymmetry}

\author[Romero, Torres, Andruchow, Anchordoqui and Link]
{Gustavo E. Romero$^{1,}$\thanks{Member of CONICET. Email:
romero@irma.iar.unlp.edu.ar}, 
Diego F. Torres$^2$, I. Andruchow$^3$, 
Luis A. Anchordoqui$^2$ \&
\newauthor 
Bennett Link$^{4,}$\thanks{Email:  blink@dante.physics.montana.edu}\\
$^1$Instituto Argentino de Radioastronom\'{\i}a,
C.C. 5, 1894, Villa, Elisa, Argentina\\
2Departamento de F\'{\i}sica, Universidad Nacional de La Plata,
C.C. 67, (1900) La Plata, Argentina\\
$^3$Facultad de Ciencias Astron\'omicas y
Geof\'{\i}sicas, Paseo del Bosque S/N, (1900)
La Plata, Argentina\\
$^4$Montana State University, Department of Physics, Bozeman,
MT 59717, USA}

\pagerange{\pageref{firstpage}--\pageref{lastpage}}

\pubyear{1999}

\begin{document}

\maketitle

\label{firstpage}

\begin{abstract} 
Based on the study of temporal asymmetry of 631 gamma ray bursts from
the BATSE 3B catalog by Link and Epstein [Ap J 466, 764 (1996)], we
identify the population of bursts whose rising times are longer than
their decays, thus showing atypical profiles. We analyse their sky
distribution, morphology, time-space clustering and other average
properties and compare them with those associated with the bulk of the
bursts. We show how most of the peculiar bursts analysed are
consistent with recent fireball models, but a fraction of bursts ($\sim
4$\% of the total sample) appear to be inconsistent.

\end{abstract}


\section{Introduction} \label{sec_introd}

During more than 25 years, the origin of gamma ray bursts (GRBs) has
been, perhaps, the deepest and most persistent problem in
astrophysics. However, with the advent of the Compton Gamma Ray
Observatory (CGRO) and its Burst and Transient Source Experiment
(BATSE) in 1991, a new phase in the research of GRBs started. In seven
years of operation, BATSE has accumulated a database of more than 2000
observations. The angular distribution of these bursts is isotropic
within the statistical limits, and the paucity of faint bursts implies
that we are seeing to near the edge of the source population (e.g.
Meegan et al. 1992, Fishman \& Meegan 1995).  Both effects, isotropy
and non-homogeneity in the distribution, strongly suggest a
cosmological origin of the phenomenon. In support of this conclusion,
absorption lines (Fe II and Mg II) in the optical counterpart of GRB
970508 have been detected with a redshift of $z=0.835$.  Along with
the absence of Lyman-$\alpha$ forest features in the spectra, these
results imply that the burst source is located at 0.835 $\le z \le $
2.3 (Metzger et al. 1997). 

The energy required to generate cosmological bursts is as high as
$10^{51}$ erg s$^{-1}$. The very short timescale observed in the time
profiles indicate an extreme compactness that implies a source
initially opaque (because of $\gamma \gamma $ pair creation) to
$\gamma$-rays.  The radiation pressure on the optically thick source
drives relativistic expansion, converting internal energy into kinetic
energy of the inflating shell. Baryonic pollution in this expanding
flow can trap the radiation until most of the initial energy has gone
into bulk motion with Lorentz factors of $\Gamma
\ge 10^2 - 10^3$. The kinetic energy, however, can be partially
converted into heat when the shell collides with the interstellar
medium or when shocks within the expanding source collide with one
another.
The randomized energy can be then radiated by
synchrotron radiation and inverse Compton scattering yielding
non-thermal bursts with timescales of seconds. This fireball scenario
has been developed by Cavallo and Rees (1978), Paczy\'nski (1986),
Goodman (1986), M\'esz\'aros and Rees (1993), M\'esz\'aros, Laguna and
Rees (1993) and others.  A comprehensive review is presented by
M\'esz\'aros (1997).

The fireball model is a robust astrophysical scenario independent of
the mechanism assumed for the original energy release. A popular
mechanism is the merger of two collapsed stars in a binary system, for
instance, two neutron stars or a neutron star and a black hole (see
Narayan, Paczy\'nski \& Piran 1992, and references threin), although
other processes have been suggested (e.g. Usov 1992, Carter 1992,
Melia and Faterzzo 1992, Woosley 1993).

One important prediction of the fireball model, as well as by
any explosive mechanism, is that individual burst
profiles should be inherently asymmetric under time reversal, with a
shorter rise time than the subsequent decay time. This is a natural
consequence of a sudden particle energy increase (e.g. produced by a
shock) and the slower radiative dissipation of the energy excess.

Time asymmetry in GRBs light curves has been discussed by several
authors (e.g.  Mitrofanov et al. 1994, Link et. al. 1993, Nemiroff et
al. 1994). In particular, Nemiroff et al. (1994) showed that in the
sample formed by those bursts with count rates greater than 1800
counts s$^{-1}$ and durations longer than 1s detected by BATSE until
1993 March 10, there is a significant asymmetry in the bursts profiles
in the sense that most bursts rise in a shorter time than they decay,
in agreement with what is expected from a general fireball model. The
most recent and complete study was made by Link and Epstein (1996).
They took 631 GRBs from the BATSE 3B catalog, including both faint and
bright bursts, and confirmed the global asymmetry in the burst 
profiles showing that about two thirds of the events display fluxes
that rise faster than the subsequent fall. About 30\% of the bursts,
however, presented a peculiar asymmetry in the temporal profiles, with
slower risings than decays.

In this paper we focus on this subsample of {\it peculiar asymmetric
bursts} (PABs), which seems at first sight to conflict with some
predictions of the simplest scenarios for fireballs.  In particular,
we shall discuss whether there are reasons to consider this subsample
of GRBs as representative of a class of sources with different
physical properties than other bursts. The structure of the paper is
as follows. In Section 2 we define the sample and present the results
of the symmetry analysis. We provide tables with the full results for
PABs in order to allow identification of specific events. In Section 3
we study the sky distribution of the sample, while in Section 4 we
investigate the level of positional coincidence (possible
repetition) that PABs show. Finally, we discuss the implications of
these results for theoretical models of GRBs.

\section{Sample and symmetry analysis}

We have studied the sample of 631 bursts from the BATSE 3B catalog
whose global symmetry properties were discussed by Link and Epstein (1996).
This sample contains both faint and bright bursts, spanning a 200-fold
range in peak flux.  PREB plus DISC data types at 64 ms time
resolution, with four energy channels, were used in the analysis.

The time asymmetry of the individual burst profiles
was examined with the skewness function introduced by Link et al. (1993)
and used in 
Link and Epstein's (1996) paper. This function is defined as

\begin{equation}
{\cal A} \equiv \frac { < \left(  t -<t> \right)^3> }
{< \left(  t -<t> \right)^2 >^{3/2}}.
\end{equation}
Here, angle brackets denote an average over the data sample, performed
as 

\begin{equation}
<g(t)> \equiv \frac{ \sum_i (c_i - c_{th}) g(t_i)}
{\sum _i (c_i - c_{th})},
\end{equation}
where $c_i$ is the measured number of counts in the $i$th bin, $t_i$
is the time of the $i$th bin and $c_{th}$ is a threshold level
defined as,

\begin{equation}
c_{th}=f(c_p -b) + b.
\end{equation}
Here $c_p$ stands for the peak (maximum) count rate, $b$ is the
background, and $f<1$ is a fraction that will be fixed for the data
set. Fixing $f$ ensures that ${\cal A}$ is calculated to the same
fraction of the peak flux relative to the background.  Larger values
of $f$ emphasize the structure of the peak over the surrounding
foothills.  The normalization of ${\cal A}$ makes it independent of 
background, duration, and amplitude. It is equal to 0 in the case of
symmetric bursts, greater than 0 for a burst whose peak rises more
quickly than it falls and smaller than 0 in the opposite case. It is
equal to 2 for an exact FRED (from {\it fast rise and exponential
decay}) and to $-2$ for an exact anti-FRED.  Four fixed values of $f$
were analyzed: $f_1$=0.1, $f_2$=0.2, $f_3$=0.5 and $f_4$=0.67.  The
only requirement for a burst to be tested in each of the four
$f$-values is that the number of bins whose $c_i$ exceeds or equals
$c_{th}$ is at least three.  Consequently, the size of the sample
differs for each choice of $f$.  The error bars in ${\cal A}$ represent
1$\sigma$ deviations, calculated by randomizing the number of counts
according to Poisson statistics and computing the variance of the
asymmetry parameter for many trials.

In Tables 1 - 4 we show the results of the skewness analysis for those
bursts that presented PAB behavior (i.e. ${\cal A} < 0$).  Each table
contains the peak flux, trigger number, burst type, value of the symmetry
parameter for each $f_i$, and a classification of the bursts profiles
according to the following scheme: S for single-peaked or spike-like
bursts, M for multiply-peaked bursts, and C for complex or chaotic
events. Regarding the burst types,  events
for which the ${\cal A}$ is negative for all $f$ are labelled
``1'', whereas events for which the errors in ${\cal A}$
allow positive ${\cal A}$ for at least one value of $f$ are
denoted ``2''. 
In Fig. 1 we show specific examples of these profiles.

We found that 91 out of 631 bursts (14.4\%) are PABs, {\sl i.e.} do not
present 
positive skewness for any $f$.\footnote{40
bursts out of these 91 are type 1.} Only 28.5\% of these bursts are
single-peaked.  The rest are multiply-peaked or complex events.
Notice that most of S-type bursts are in Table 4.  This is consistent
with the analysis technique: a fast burst, typically lasting a couple
of seconds, will have few points above the higher cut-offs and then,
data for ${\cal A}_{f_2,f_3,f_4}$ will not be computed.

As we can see from Tables 1 - 4 as well as from Fig. 1, PABs exhibit a
variety of temporal morphologies.  If all of these events are
produced by a single mechanism, then there should be a very wide range
of boundary and initial conditions in the sources in order to generate
such a plurality of profiles. With the aim of searching for
differences between PABs and the more common bursts, we have computed
average values of the hardness ratio and durations of type 1 PABs.
These values are compared with similar estimates for those bursts with
${\cal A} > 0$ at all levels in Table 5.  Due to the small number of
bursts and the variety in their features, dispersions are so large
that no conclusions can be drawn. However, in the 
light of current data, it is clear that no significant correlation
is found between hardness ratio, or duration, with burst morphology.

\section{Sky distribution}

One of the most important results of BATSE is the discovery of that
GRBs are isotropically distributed on the sky
(see, however, Balazs et al. 1998).
With the recent
detection of high-redshifted absorption lines in the optical
counterparts of individual bursts (Djorgovski et al. 1997, Metzger et
al. 1997, van Paradjis et al. 1997) Galactic  models appear to be
finally ruled out. However, one could ask whether the distribution of
PABs exhibits the same level of isotropy than that of the whole
sample. It could be the case, for instance, that PABs have a different
origin than other GRBs, and consequently, display a distinct 
distribution on the sky 
(e.g. there could be a statistically significant concentration
of PABs in the supergalactic plane or within any superstructure).

In order to quantify the isotropy we followed the method developed by
Briggs (1993).  The dipole moment toward the Galactic  center is $<\cos
\theta > $, the mean of $\cos \theta_i$, where $\theta_i$ is the angle
between the $i$th burst and the Galactic  center. An excessively large
value of $<\cos \theta >$ indicates a significant dipole moment towards
the Galactic  center. The quantity $<\sin^2 b -1/3 >$ tests for a
concentration in the Galactic  plane or in the Galactic poles. The
expected mean values of the two statistics are zero for an isotropic
distribution and, if they are asymptotically gaussian distributed, 
i.e. if for a large number of bursts in the sample ($N$) 
they are gaussian distributed, 
the variances $\sigma^2$ are $1/3N$ and $4/45 N$
respectively.

Briggs et al. (1996) noted that because the CGRO is in a low-Earth
orbit, about one-third of the sky is blocked by the Earth causing a
portion of the Galactic equator to be observed about 20\% less than the
poles. An additional effect is different exposure times between the
Galactic south and north poles. These effects must be taken into
account when computing the expected values of the statistics (Briggs
et al. 1996). Location errors on particular bursts, however, have no
impact on the isotropy characteristics because they are small compared
with the large scale of anisotropies we are testing against.

Table 6 shows that the distribution of all PABs (Fig. 2) is consistent
with perfect isotropy. The same is 
true for sub-samples
of PABs. Some entries in Table 6 show small deviations from isotropy
(quadrupole).  However, the small number of events make the asymptotic
gaussian distribution no longer valid, and one should compare with the
study of Briggs et al. (1996) (see their Fig. 4a and b).  
Comparing with the values of $\sigma$ that arise from the previous
cited figures of the Briggs et al. work, we find,
consequently, that there is no detectable anisotropy in the sky
distribution of PABs and 
we see that the 1$\sigma$ deviation from isotropy
contains the values of all entries in Table 6.

\section{Time-space clustering}

Several time-space clustering analysis of different GRB-samples have
come to contradictory conclusions about whether some GRBs repeat or not
(e.g. Quashnock \& Lamb 1993, Narayan \& Piran 1993, Wang \&
Lingenfelter 1995, Petrosian \& Efron 1995, Meegan et al. 1995). The
most complete study on the subject until now, carried out by Tegmark
et al. (1996), is based on the analysis of the angular power spectrum
of 1120 bursts from BATSE 3B catalog. These authors found that the
number of bursts that can be labelled as repeaters (considering just
one repetition) is not larger than 5\% at 99\% confidence.
The recent study of by Gorosabel et al. (1998), which
combined data from different satellites, shows that at most 15.8\% of
the events detected by WATCH recur in the BATSE sample (at 94\%
confidence level). Despite the discussion in the literature, it seems
clear that only a small fraction of the total number of GRBs
could repeat over timescales of up to a few years.

However, if PABs have a different physical origin than other bursts,
this subclass of bursts might exhibit time-space clustering. In fact,
we find that 48 out of 91 PABs (52.7\%) have companions within their
location error boxes in the sample of 631 bursts. If we consider just
bursts separated by less than 4$^o$, we find 40 possible repeaters
(44\% of the PAB-subsample); typically, the separation is about
2.5$^o$.

To estimate the statistical significance of these results, we have
made a numerical study as follows. We have simulated 1500 sets of 91
random positions for PABs. In order to do this, we have made rotations
on the celestial sphere sending a particular PAB with coordinates
$(l,b)$ to a new position $(l',b')$, which is obtained from the
previous by seting $l'=l+ R_1 \, 360^o$ and $b'=b + R_2 \, 90^o$,
and using appropriate spherical boundary conditions. Here, $R_1$ and
$R_2$ are different random numbers (between 0 and 1) which never
repeat. Doing this for each event we get a new set of simulated
PAB-positions. For this set we then compute the positional coincidence
level with respect to the fixed $631-91$ GRBs coordinates.  As in the
real case, we shall assign a positional coincidence when two or more
bursts are separated by less than 4$^o$.  After making 1500 operations
of this type (a larger number of simulations does not significantly
modify the results) we can obtain the mean value of the expected
number of positional coincidences and its $\sigma$. We obtain that for
91 GRBs, the average level of positional coincidences is 42.9 $\pm$
4.7, which is entirely compatible with the observed result for PABs
within 1$\sigma$.

We have repeated the process for the subset of 26 single-peaked PABs
(those denoted by an S in Tables 1 - 4). These events represent about
4\% of the whole sample and about 28.5\% of the PAB subsample. 15 out
of 26 bursts of this kind ($\sim$60\%) present companions within error boxes
of less than 4 degrees.  
We find that the average 
simulated positional coincidence level is 13.3 $\pm$ 2.5.
That is, the real coincidence level is also compatible with the random
result to within 1$\sigma$ and no particular association appears obvious.

If we now take positional coincidences separated
by less than 1$^o$, we find that 3 out of 26 single-peaked PABs have 
companions. Repeating the simulations in this case yields an expected chance
association of 1 $\pm$ 1 events. This means that the real positional
coincidence is only compatible with the random one to within 2$\sigma$.
The number of events is of course too scarce to draw any conclusion, but if
this is confirmed in a larger sample it would entail an excess of 3.8\%
repetitions above 
the result expected from chance associations (something compatible with
already mentioned 
Tegmark et al.'s analyses).

As we shall see in the next section, spikes with peculiar asymmetry
present problems for their interpretation within the standard
fireballs models.

\section{PABs and the fireball models}

As mentioned in the introduction, GRBs profiles with ${\cal A} <0$ are
not expected from the simplest versions of the fireball model (i.e. a
single expanding shell that acts as a gamma ray emitter during a brief
time at some fixed radius from the central site of the explosion, e.g.
Fenimore et al. 1996).  However, one of the distinctive features of
the fireball scenario is that the same basic mechanism can generate a
variety of time profiles for different initial and boundary
conditions. We now discuss whether these changes can provide the main
types of PABs observed in the sample.

In Fig. 3 we show the profile of BATSE trigger \#2450, which has
negative skewness function for all values of $f$ (see Table 1). This
is a typical multiply-peaked burst, with a precursor at $t=0$ and a
series of peaks of increasing height that start about 35 s after the
first signal. Individual peaks, when analysed with appropriate
$f$, give ${\cal A}>0$.  Events of this kind can be understood as the
effect of a mild baryon loaded fireball (M\'esz\'aros \& Rees 1993).
Even a small baryon contamination ($M_b \ge 10^{-9} M_\odot$) of the
expanding pair-photon fireball is enough to trap the $\gamma$-rays
until most of the initial energy is transformed into kinetic energy of
the baryons. The fireball expands by radiation pressure and becomes
optically thin to Thomson scattering when the optical depth drops
below unity at a radius given by (M\'esz\'aros \& Rees 1993),

\begin{equation}
r_p \sim 0.6 \times 10^{15} \theta^{-1} E_{51}^{1/2} \eta^{-1/2}\;\; {\rm cm},
\end{equation}
where $\theta$ takes into account the possibility of channeling of the
flow ($\theta \sim 1$ corresponds to spherical symmetry), $E_{51}$ is
the original energy release ($e^\pm, \gamma$) in units of 10$^{51}$
erg, and $\eta = E_0/M_0 c^2$ is the initial radiation to the rest
mass energy ratio in the
fireball. At $r=r_p$, the $\gamma$-rays still trapped in the fireball
can escape producing a burst (Cavallo \& Rees 1978, Pacz\'nski 1986,
Goodman 1986). As shown by M\'esz\'aros \& Rees (1993), this burst
should be rather weak, with an observed energy in gamma rays of,
approximately, 

\begin{equation}
E_p^{obs} \sim 7 \times 10^{47} \theta^{1/3} E_{51}^{1/2} \eta_3^{11/6} \;\;
{\rm erg},
\end{equation}
where $\eta_3=10^{-3} \eta$. This
prompt, small burst will form a precursor that can last a few seconds.
When the expanding relativistic shell collides with the interstellar
medium, a shock wave is formed and the gas in the post-shock region
is heated up to thermal Lorentz factors of $\gamma \sim \eta$,
reconverting the kinetic energy of the shell into thermal energy of
the particles. The thermal energy is radiated through synchrotron and
inverse Compton processes at MeV to GeV energies. A non-uniform
ambient medium can naturally lead to a multiply-peaked burst (e.g.
Fenimore et al. 1996).  Events of this class will have ${\cal A}<0$,
as in the case of trigger
\#2450, due to the effect of the prompt precursor. Hence, these ${\cal
A}<0$ events can be explained within the fireball model.

In other cases, the precursor can remain undetected but a multiply-peaked
PAB can arise from internal shocks in bursts with several shells
with different Lorentz factors
(e.g. Kobayashi et al. 1997, Daigne \& Mochkovitch 1998).
In the simulations carried out
by Kobayashi et al. (1997), bursts with negative skewness 
can be produced through multiple shell collisions
(e.g. see Fig. 2f of their work).

Complex bursts, as the one shown in Fig. 1d, could be the result of
instabilities on the expanding shell surface once it shocks the
interstellar medium.  Hydromagnetic instabilities in the contact discontinuity 
can lead to local variations in the fields and the flow's Lorentz factor,
yielding very rapid changes in the time profiles
(e.g. Daigne \& Mochkovitch 1998). The resulting global morphology
could resemble that seen in some bursts with negative
skewness, such as \#2240.

Single-peaked bursts with ${\cal A}<0$, however, appear to be more
difficult to explain with the fireball model.  The main problem is
that a single spike with slower rising than falling cannot be
generated through dissipative shocks. In Fig. 4 we show BATSE trigger
\#444 (see also Table 3 and Fig. 1a). We have attempted to fit this event
with the multiple shell
model developed by Kobayashi et al. (1997). The $\gamma$-ray emission
is produced when a shock results from the collision of two shells with
different velocities. The randomized kinetic
energy is then radiated through synchrotron and inverse Compton processes.
Notice that the better the fit for the rising profile, the worse the
model describes the fall.  This is a straightforward consequence of
the fact that cooling times are longer than particle acceleration
times at the shock.

To better understand the meaning of the theoretical curves
in Fig. 4 we recall the predicted luminosity in the case of a two
shell interaction (Kobayashi et al. 1997),
\begin{equation}     
{\cal L}(t)  \propto \left\{
\begin{array}{l}
1- ( 1 + 2 \gamma_m^2 c t /R)^{-2},  
            \hspace{0.5cm} 0<t<\delta t_e /2 \gamma_m^2  \\
( 1 + (2 \gamma_m^2  t - \delta t_e) c /R)^{-2}
                        -( 1 + 2 \gamma_m^2 c t /R)^{-2 }, \\
\hspace{0.8cm}                        t>\delta t_e /2 \gamma_m^2
\end{array}                  
\right.
\end{equation}
where $\gamma_m$ is the Lorentz factor of the merged shell (depending
on the Lorentz factor and mass of each colliding shell), $\delta
t_e / 2 \gamma_m^2$ is the time at which the burst reach its maximum,
and $R$ is the radius at which the collision takes place.
Observational data of a given burst, its height and duration up to the
maximum in the number of counts, allow a parameterization of ${\cal
L}(t)$ with 

\begin{equation}
B=\frac{2 \gamma_m^2 c }{R}  .
\end{equation}                     
The shape of the pulse is asymmetric with a fast rise and a slower
decline unlike a spike event with ${\cal A}<0$. Attempts to fit such
a burst using eq. (6) are shown in Fig. 4.
 
Spike-like bursts with ${\cal A}<0$ are predicted, however, in some
extrinsic models for GRBs. Torres et al. (1998a,b) have shown that
microlensing effects produced upon the core of high redshifted AGNs by
compact extragalactic  objects which violate the weak energy condition
at a
macroscopic level would yield GRB-like lightcurves with spike-type
profiles and negative skewness function. A similar burst
with ${\cal A}>0$ should be observed from several months up to a few
years later in the same position of the sky, provided the lens has an
absolute mass of the order of $1M_\odot$. If this interpretation turns
out to be correct, it could explain not just S-type PABs but also any
apparent excess of positional coincidences among these bursts at a level
compatible with current constraints on repetition over the whole sample. 
It would appear that the
small group of spike bursts with negative skewness deserve further
study. 

\section{Conclusions}

GRBs exhibit a very rich variety of temporal profiles. Most of them have
highly variable structure over timescales significantly shorter than
the overall duration of the event. The study of burst morphology by
Link and Epstein (1996) shows that a significant fraction of bursts
($\sim 1/3$) have time histories in which the flux rises more rapidly
than it decays (PABs). Here we have argued that most PABs can be
accommodated by fireball models.  Isotropy and other average features,
common to the bulk of observed bursts, 
are shared by PABs. But there is, however, a subclass of PABs, those which
consist of a single, prominent
peak with negative skewness, that appear to be inconsistent with the
fireball mechanism. These events represent $\sim 4$ \% of the total sample
and certainly merit further research in order to clarify their nature. 

\section*{Acknowledgments}
This research has made use of data obtained 
through the High Energy Astrophysics Science 
Archive Research Center, provided by the NASA/Goddard
Space Flight Center 
and also of 
NASA/IPAC Extragalactic Database, which is operated by the Jet Propulsion
Laboratory, California Institute of Technology, under contract with
NASA.
Our work has been supported by the Argentine agencies CONICET 
(D.F.T. and G.E.R - under grant PIP N$^o$ 0430/98 -), 
ANPCT (G.E.R.), and FOMEC (L.A.A.).
We acknowledge G. Bosch for his help in producing Fig. 2 and S. Grigera
for an enlightening discussion on numerical issues.

{}

\clearpage

\begin{table*}
\label{1}                                            
\centering
\begin{minipage}{180mm}
\caption{GRBs with negative skewness function for all four 
cut-off parameters. Type 1 indicates bursts whose errors in ${\cal A}$ are
less than the modulus of ${\cal A}$ itself and type 2 indicates 
cases in which at least one of the computed errors in ${\cal A}$
exceeds the modulus of ${\cal A}$. S, M, and C stand for 
single, multiply-peaked and complex temporal profiles.}

\begin{tabular}{ll|clllll}

Peak flux &  \# & Type & ${\cal A}_{f_1}$ &       
${\cal A}_{f_2}$ & ${\cal A}_{f_3}$ & ${\cal A}_{f_4}$  & Profile \\
\hline

20.23800 & 1663 & 1 & -9.55E-02 $\pm$ 1.92E-03 & -0.14 $\pm$ 2.56E-03 &
           -0.17 $\pm$ 1.13E-02 & -0.15 $\pm$ 5.21E-02 & M \\

20.13300 & 219 &1 & -0.20 $\pm$ 9.21E-03 & -0.20 $\pm$ 8.48E-03 &
           -1.1 $\pm$ 4.06E-02 & -0.15 $\pm$ 6.84E-02  & M   \\

13.78700 & 1122 & 1 & -0.30 $\pm$ 7.69E-02 & -0.72 $\pm$ 0.10 &
           -0.21 $\pm$ 5.82E-02 & -0.72 $\pm$ 0.15  & M  \\

8.79800 & 2834 & 1 & -4.42E-02 $\pm$ 2.86E-02 & -6.20E-02 $\pm$ 4.02E-02 &
           -8.69E-02 $\pm$ 5.62E-02 & -0.13 $\pm$ 8.24E-02  & M\\

8.27700 & 2450 & 1 & -0.42 $\pm$ 9.27E-03 & -7.51E-02 $\pm$ 3.24E-03 &
           -0.25 $\pm$ 6.46E-03 & -0.65 $\pm$ 9.21E-02 & M \\

7.13100 & 2852 & 2 & -0.15 $\pm$ 8.97E-03 & -1.17E-02 $\pm$ 4.13E-02 &
           -5.91E-02 $\pm$ 5.27E-02 & -0.16 $\pm$ 0.46 & M\\

6.71400 & 2436 & 1 & -0.71 $\pm$ 8.86E-02 & -2.87E-02 $\pm$ 5.27E-03 &
           -0.27 $\pm$ 7.69E-02 & -0.48 $\pm$ 1.95E-02 & M  \\

5.69900 & 1974 & 1 & -9.82E-02 $\pm$ 8.65E-03 & -0.15 $\pm$ 8.49E-03 &
           -0.36 $\pm$ 0.16 & -4.31E-02 $\pm$ 3.76E-02 & M \\

5.24900 & 179 & 1 & -0.62 $\pm$ 9.11E-02 & -0.73 $\pm$ 0.24 &
           -0.18 $\pm$ 5.49E-02 & -0.27 $\pm$ 0.12   & S\\

4.53800 & 222 & 2 & -0.82 $\pm$ 0.22 & -0.18 $\pm$ 3.41E-02 &
           -0.23 $\pm$ 7.24E-02 & -0.11 $\pm$ 0.15 & M\\

3.02000 & 2922 & 2 & -0.35 $\pm$ 1.80E-02 & -0.53 $\pm$ 0.14 &
           -0.13 $\pm$ 0.11 & -0.32 $\pm$ 0.43 & C\\

2.79600 & 906 & 2 & -9.97E-02 $\pm$ 6.91E-02 & -0.11 $\pm$ 5.89E-02 &
           -0.22 $\pm$ 0.16 & -0.18 $\pm$ 0.19 & S\\

2.34200 & 2428 & 2 & -0.14 $\pm$ 7.57E-03 & -8.68E-02 $\pm$ 8.95E-02 &
           -0.13 $\pm$ 2.24E-02 & -0.52 $\pm$ 4.63E-02  & M \\

2.06600 & 2476 & 1 & -2.28E-02 $\pm$ 2.35E-03 & -5.84E-03 $\pm$ 4.12E-03 &
           -0.11 $\pm$ 3.35E-02 & -0.20 $\pm$ 3.06E-02 & M\\

1.71900 & 353 & 2& -0.23 $\pm$ 2.14E-02 & -0.13 $\pm$ 3.25E-02 &
           -9.92E-02 $\pm$ 1.86E-02 & -0.12 $\pm$ 0.28 & S\\

1.64800 & 2074 & 1 & -9.33E-02 $\pm$ 4.34E-02 & -0.19 $\pm$ 9.48E-02 &
           -0.31 $\pm$ 0.11 & -0.25 $\pm$ 0.20 & M\\

1.54500 & 254 & 2 & -3.68E-02 $\pm$ 7.61E-02 & -6.36E-02 $\pm$ 8.59E-02 &
           -0.15 $\pm$ 0.19 & -0.19 $\pm$ 0.36 & S \\

1.44900 & 1197 & 2 & -0.27 $\pm$ 4.82E-02 & -0.29 $\pm$ 0.14 &
           -0.17 $\pm$ 17 & -0.25 $\pm$ 0.27  & C \\

1.37400 & 1413 & 2 & -0.18 $\pm$ 0.20 & -0.29 $\pm$ 0.24 &
           -2.68E-02 $\pm$ 0.30 & -0.14 $\pm$ 0.27  & M\\

1.35300 & 2610 & 1 & -0.31 $\pm$ 3.31E-02 & -0.19 $\pm$ 3.28E-02 &
           -0.87 $\pm$ 6.81E-02 & -0.39 $\pm$ 0.21 & C\\

1.11400 & 752 & 2 & -0.19 $\pm$ 0.14 & -0.24 $\pm$ 0.13 &
           -0.61 $\pm$ 0.29 & -0.32 $\pm$ 0.34 & S\\

1.11000 & 2828 & 2 & -7.11E-02 $\pm$ 0.29 & -9.19E-02 $\pm$ 9.92E-02 &
           -0.14 $\pm$ 0.16 & -0.15 $\pm$ 0.27   & C\\

0.84500 & 2448 & 2 & -9.77E-02 $\pm$ 1.87E-02 & -0.11 $\pm$ 2.28E-02 &
           -7.29E-03 $\pm$ 7.45E-02 & -0.11 $\pm$ 8.27E-02  & C\\

0.77600 & 2495 & 1 & -2.43E-02 $\pm$ 1.23E-02 & -4.93E-02 $\pm$ 1.22E-02 &
           -0.15 $\pm$ 0.14 & -1.20 $\pm$ 0.68 & C\\

0.66100 & 690 & 2 & -0.63 $\pm$ 0.21 & -0.35 $\pm$ 0..32 &
           -0.19 $\pm$ 0.37 & -0.30 $\pm$ 0.46    & C\\

0.59400 & 2442 & 2 & -2.36E-02 $\pm$ 3.01E-02 & -1.03E-02 $\pm$ 0.11 &
           -0.13 $\pm$ 0.17 & -0.22 $\pm$ 0.14 & C\\

\hline

\end{tabular}
\end{minipage}
\end{table*}

\begin{table*}
\label{2}                                            
\centering
\begin{minipage}{160mm}
\caption{GRBs with 
negative skewness for three thresholds $f$. For $f_4$, these bursts
did not fulfill the criteria for the assignment of an
${\cal A}$ value.  Burst type and capital
letters have the same meaning as in Table 1.}

\begin{tabular}{ll|cllll}

Peak flux &\# & Type & ${\cal A}_{f_1}$ &       
${\cal A}_{f_2}$ & ${\cal A}_{f_3}$  & Profile\\
\hline

12.45800 & 1440 & 1 & -7.47E-02 $\pm$ 8.56E-03 & -0.15 $\pm$ 8.27E-03 &
           -0.35 $\pm$ 0.17  & M\\

8.73700 & 2799 & 1 & -0.78 $\pm$ 6.23E-03 & -1.1 $\pm$ 0.23 &
           -0.29 $\pm$ 8.58E-02 & M \\

2.37600 & 2861& 2 & -2.27E-02 $\pm$ 7.87E-02 & -0.12 $\pm$ 0.12 &
           -7.99E-02 $\pm$ 0.27 & M \\

2.35100 & 2795 & 1& -0.13 $\pm$ 1.62E-02 & -0.14 $\pm$ 1.92E-02 &
           -0.14 $\pm$ 5.31E-02  & S\\

2.00500 & 1359 & 2 & -0.11 $\pm$ 0.14 & -0.14 $\pm$ 8.14E-02 &
           -0.76 $\pm$ 0.27  & S\\

1.93900 & 1154 & 2 & -6.24E-02 $\pm$ 6.27E-02 & -8.17E-02 $\pm$ 6.22E-02 &
           -0.44 $\pm$ 0.24  & S\\

1.23000 & 1924 & 1 & -0.22 $\pm$ 0.12 & -0.16 $\pm$ 0.11 &
           -0.17 $\pm$ 0.13  & S\\

1.22600 & 1611 & 2 & -0.73 $\pm$ 0.34 & -0.20 $\pm$ 0.29 &
           -0.56 $\pm$ 0.26  & C\\

1.20400 & 3012 & 2 & -2.87E-02 $\pm$ 0.12 & -0.15 $\pm$ 0.16 &
           -9.32E-02 $\pm$ 0.22 & M \\

1.20200 & 3080 & 2 & -0.90 $\pm$ 0.36 & -0.17 $\pm$ 0.20 &
           -0.57 $\pm$ 0.25  & C\\

1.19100 & 2040 & 2 & -4.77E-02 $\pm$ 0.16 & -5.50E-02 $\pm$ 0.18 &
           -8.62E-02 $\pm$ 0.25 & C\\

1.05000 & 3017 & 2 & -6.30E-02 $\pm$ 0.10 & -0.17 $\pm$ 0.17 &
           -0.10 $\pm$ 0.19  & C\\

0.88000 & 2204 & 2 & -9.25E-02 $\pm$ 0.28 & -0.41 $\pm$ 0.31 &
           -0.29 $\pm$ 0.22  & C\\

0.81300 & 2240 & 2 & -0.18 $\pm$ 3.84E-02 & -0.20 $\pm$ 5.62E-02 &
           -0.10 $\pm$ 0.12  & C\\

0.74000 & 114 & 2 & -0.13 $\pm$ 0.12 & -0.11 $\pm$ 0.28 &
           -0.38 $\pm$ 0.45 & C\\

0.67800 & 2434 & 1 & -0.31 $\pm$ 2.74E-02 & -0.37 $\pm$ 3.15E-02 &
           -1.0 $\pm$ 0.24  & S\\

0.54600 & 2727 & 2 & -7.14E-03 $\pm$ 0.11 & -1.43E-02 $\pm$ 8.57E-02 &
           -0.33 $\pm$ 0.50  & C\\

0.34200 & 1435 & 2 & -0.15 $\pm$ 0.25 & -0.48 $\pm$ 0.23 &
           -0.65 $\pm$ 0.26  & C\\
             
\hline

\end{tabular}
\end{minipage}
\end{table*}

\begin{table*}
\label{3}
\centering
\begin{minipage}{160mm}
\caption{GRBs with ${\cal A}<0$ for two
thresholds. }

\begin{tabular}{ll|clll}

Peak flux & \# & Type & ${\cal A}_{f_1}$ &       
${\cal A}_{f_2}$ & Profile\\
\hline

28.55900 & 444 &1 & -0.23 $\pm$ 1.66E-02 & -0.45 $\pm$ 3.51E-02  & S\\

4.25500 & 2993&2 & -0.53 $\pm$ 0.58 & -9.09E-02 $\pm$ 5.33E-02 & M\\

3.26900 & 2041 &1 & -0.61 $\pm$ 8.86E-02 & -1.1 $\pm$ 0.36  & M\\

2.22800 & 2220 &2 & -0.44 $\pm$ 0.46 & -1.66E-02 $\pm$ 0.15  & S\\

1.89300 & 2201 &1 & -0.26 $\pm$ 0.15 & -0.32 $\pm$ 7.65E-02  & S\\

1.85500 & 2788 &2 & -4.68E-02 $\pm$ 0.10 & -5.79E-02 $\pm$ 1.80E-02  & S\\

1.22900 & 1142 & 1 & -0.83 $\pm$ 0.39 & -1.1 $\pm$ 0.42 & S\\

1.20400 & 1968 & 1 & -0.26 $\pm$ 0.19 & -0.29 $\pm$ 0.14 & S \\
              
1.07700 & 171 & 2 & -7.18E-02 $\pm$ 0.26 & -9.49E-02 $\pm$ 0.27  & C\\

1.06800 & 2529 &1 & -0.40 $\pm$ 5.15E-02 & -0.35 $\pm$ 0.11  & M\\

1.06000 & 2853 & 2 & -3.79E-02 $\pm$ 9.67E-02 & -8.85E-02 $\pm$ 0.14 & C\\

0.96700 & 237 & 2 & -0.11 $\pm$ 0.57 & -0.38 $\pm$ 0.46 & C\\

0.97900 & 2800 & 1 & -0.21 $\pm$ 0.12 & -0.27 $\pm$ 0.17  & C\\

0.82500 & 2290 & 1 & -0.47 $\pm$ 4.08E-02 & -0.52 $\pm$ 4.27E-02  & C\\

0.80800 & 2129 & 2 & -2.30E-02 $\pm$ 0.20 & -0.16 $\pm$ 0.23 & C \\

0.78200 & 1192 & 2 & -2.31E-02 $\pm$ 0.13 & -3.93E-02 $\pm$ 0.20& C  \\

0.78800 & 2857 & 1 & -0.25 $\pm$ 0.16 & -0.32 $\pm$ 0.18 & C\\

0.76600 & 1430 & 2 & -8.19E-02 $\pm$ 0.26 & -0.10 $\pm$ 0.24 & C \\

0.71600 & 1404 & 2 & -1.62E-02 $\pm$ 0.17 & -1.85E-02 $\pm$ 0.22 & C \\

0.71300 & 1110 & 2 & -0.11 $\pm$ 0.14 & -1.73E-02 $\pm$ 0.14  & C\\

0.70800 & 204 & 2 & -0.25 $\pm$ 0.10 & -0.11 $\pm$ 0.14  & M\\

0.70400 & 2776 & 1 & -0.23 $\pm$ 8.03E-02 & -0.22 $\pm$ 0.10 & C \\

0..65400 & 2996 & 2 & -0.17 $\pm$ 0.23 & -4.76E-02 $\pm$ 0.48  & C\\

0.65100 & 2725 & 1 & -0.25 $\pm$ 8.81-02 & -0.40 $\pm$ 0.19  & C\\

0.61800 & 1655 & 2 & -7.66E-02 $\pm$ 0.31 & -0.37 $\pm$ 0.28 & C \\

0.57600 & 2750 & 2 & -4.14E-02 $\pm$ 0.13 & -0.11 $\pm$ 5.83E-02  & C\\

0.56300 & 2230 & 2 & -0.10 $\pm$ 0.20 & -0.14 $\pm$ 0.18 & C \\

0.56100 & 2069 & 2 & -9.99E-02 $\pm$ 0.15 & -0.37 $\pm$ 0.20  & C\\

0.55200 & 1382 & 2 & -0.14 $\pm$ 0.18 & -0.28 $\pm$ 0.16  & C\\

0.54500 & 2900 & 2 & -0.33 $\pm$ 0.24 & -0.12 $\pm$ 0.30  & C\\

0.53400 & 1465 & 2 & -8.62E-02 $\pm$ 0.23 & -2 $\pm$ 0.34 & C \\

0.49800 & 2233 & 1 & -0.48 $\pm$ 0.38 & -0.51 $\pm$ 0.13 & C \\

0.47500 & 559 &   1 & -0.49 $\pm$ 0.31 & -0.31 $\pm$ 0.24 & C\\

0.44000 & 1120 & 2 & -1.27E-02 $\pm$ 0.19 & -7.57E-03 $\pm$ 0.22  & C\\

\hline

\end{tabular}
\end{minipage}
\end{table*}

\begin{table*}
\label{4}
\centering
\begin{minipage}{160mm}
\caption{GRBs with ${\cal A}<0$ for one
threshold. }
\begin{tabular}{ll|cll}
Peak flux & \# & Type & ${\cal A}_{f_1}$ & Profile\\ \hline

8.76100 & 2978 & 1 & -1.60 $\pm$ 0.95  & S\\

5.57300 & 551 & 1 & -9.78E-02 $\pm$ 5.33E-02 & S \\

4.68100 & 1851 & 1 & -1.1 $\pm$ 0.44  & S\\

4.08600 & 2995 & 1 & -0.83 $\pm$ 0.25 & S \\

3.43700 & 2918 & 1 & -0.24 $\pm$ 0.20 & S \\

2.97900 & 297 &  1 & -0.11 $\pm$ 3.65E-02  & M\\

1.92600 & 2161 & 1 & -1.0 $\pm$ 0.63  & S\\

1.79100 & 2846 & 1 & -0.36 $\pm$ 0.18& S \\

1.67000 & 1461 & 1 & -0.95 $\pm$ 0.41& S \\

1.13300 & 2163 & 2 & -0.13 $\pm$ 0.13& S  \\

1.03200 & 2823 & 2 & -0.32 $\pm$ 0.34& S  \\

0.69500 & 2508 & 1 & -1.6 $\pm$ 0.74  & C\\

0.64800 & 2437 & 1 & -0.17 $\pm$ 4.50E-02  & C\\

\hline

\end{tabular}
\end{minipage}
\end{table*}

\begin{table*}
\label{16}
\centering
\begin{minipage}{160mm}
\caption{Mean values of hardness ratio and temporal durations
for type 1 PABs  compared with 
usual ${\cal A}>0$ bursts. All bursts have ${\cal A}_{f_i}$ computed
for all four $f_i$ analysed. The PAB group has $\sim$26 bursts, whereas
the ${\cal A}>0$ set comprises $\sim$100 of the total sample. }

\begin{tabular}{lcccc}

             & $<$Hard. Ratio$>$ & $\sigma$ & $<T_{90}>$ & $\sigma$ \\\hline
type 1= PABs & 3.6 & 2.3 & 35 & 42  \\ 
type 1= GRBs &  3.2 & 1.5 & 49& 50  \\ 
\hline

\end{tabular}
\end{minipage}
\end{table*}

\begin{table*}
\label{17}
\centering
\begin{minipage}{160mm}
\caption{Isotropy characteristics. The 
expected values of the statistics (corrected for BATSE exposure) are
-0.013 for $<\cos \theta>$ and -0.005 for $<\sin^2 b -1/3>$.}

\begin{tabular}{lcc|cc}
Sample &   $<\cos \theta >$ &  $\sigma =0.99 \sqrt{1/3N_B}$  
&   $<\sin^2 b -1/3>$ & $\sigma=0.99 \sqrt{4/45 N_B}$ 
\\ \hline
1+2-types=91 GRBs & -0.034 & 0.059 &  -0.023 & 0.030 \\ 
1-type=41 GRBs & 0.003 & 0.089 & -0.045 & 0.046 \\
2-type=49 GRBs & -0.065 & 0.082 & -0.005 & 0.042 \\
1-type Table 1=12 GRBs& -0.120 & 0.165 & -0.052 & 0.085 \\
1-type Table 3=13 GRBs & 0.072 & 0.158 & -0.092 & 0.081 \\
1-type Table 1+ 2=17 GRBs & 0.030 & 0.138 & -0.104 & 0.071 \\
1-type Table 1 + 2 + 3=30 GRBs & -0.014 & 0.104 & -0.098 & 0.053 \\
single-peaked =26 GRBs & 0.015 & 0.110 & 0.015 & 0.057 \\
Multiply-peaked and complex =65 GRBs & -0.045 & 0.071 & -0.044 & 0.036 \\
\hline
\end{tabular}
\end{minipage}
\end{table*}

\clearpage

\begin{figure}   
\caption{Examples of bursts with negative skewness. From top
to bottom and left to right, Fig. 1a and 1b are single-peaked bursts.
Fig. 1c is an example of a multiply-peaked burst with 
${\cal A}<0$ and Fig. 1d is a complex peculiar asymmetric
burst. For details on the particular ${\cal A}$-values see Tables 1 - 4.}
\label{fig1}
\end{figure}

\begin{figure}   
\caption{Spatial position (Galactic coordinates, Aitoff projection)
of the PABs presented in Tables 1 - 4. The sky distribution is consistent with 
perfect isotropy.}
\label{fig2} 
\end{figure}
\vspace{2cm}

\begin{figure}   
\caption{An example of a burst with negative overall skewness, but
positive skewness for the subpeaks. A burst of this type could be made up of
a precursor associated with baryonic pollution, followed by a series of peaks
generated in the collision of a relativistic shell with the
interstellar medium.}
\label{fig3} 
\end{figure}

\vspace{2cm}

\begin{figure}   
\caption{An illustration of the inconsistency between the fireball
model and spike bursts with negative skewness. Shown is the spike
burst of Fig. 1a (see also Table 3). The bar plot stands for  
the observed number of counts.
We also show  
several profiles of the light curves predicted by the multiple shell
model of Kobayashi et al. (1997) corresponding to 
$B= 0.1, 5$ and 15 (see text). 
In general, the better the theoretical
curves fit the rising curve, the worse they fit the falling portion.}
\label{fig4} 
\end{figure}

\label{lastpage}

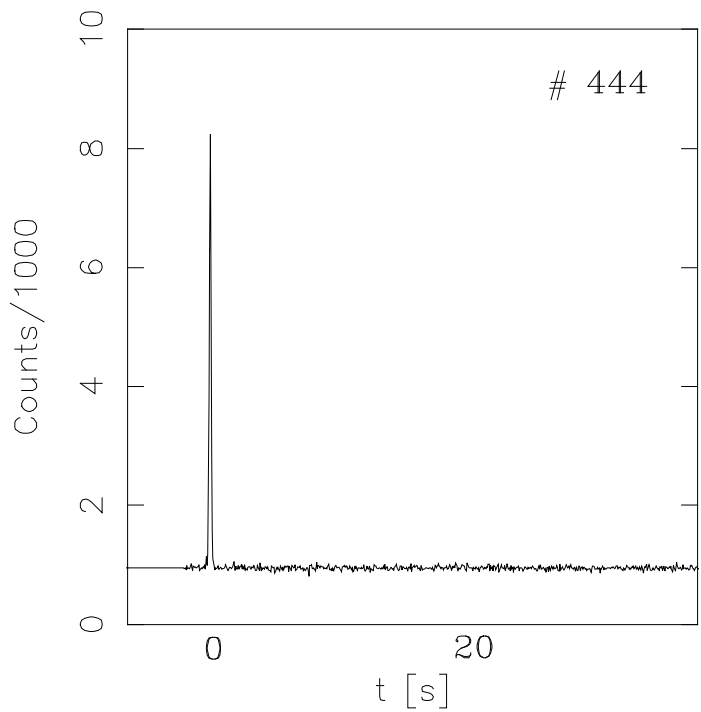
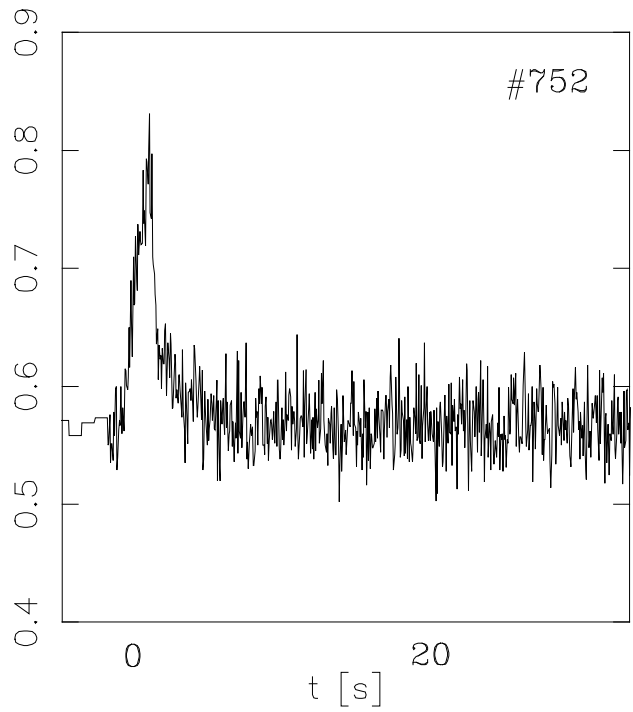
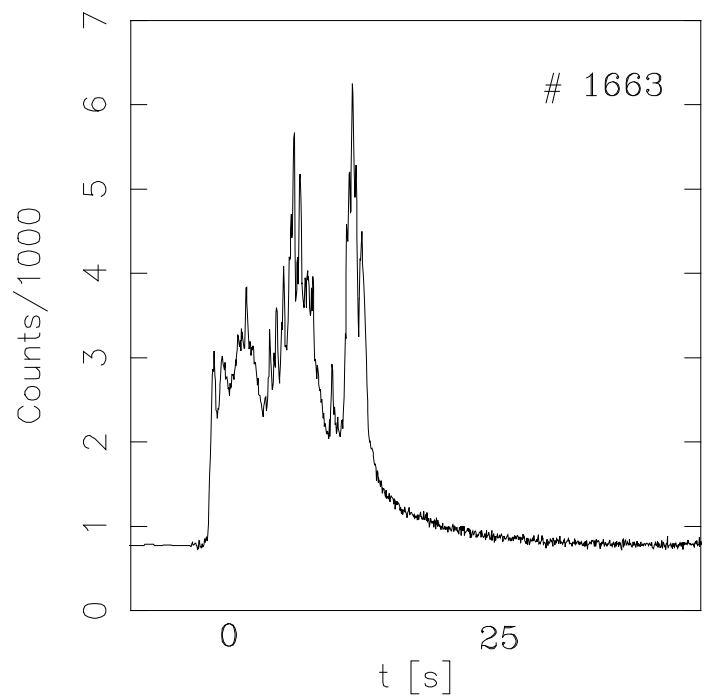
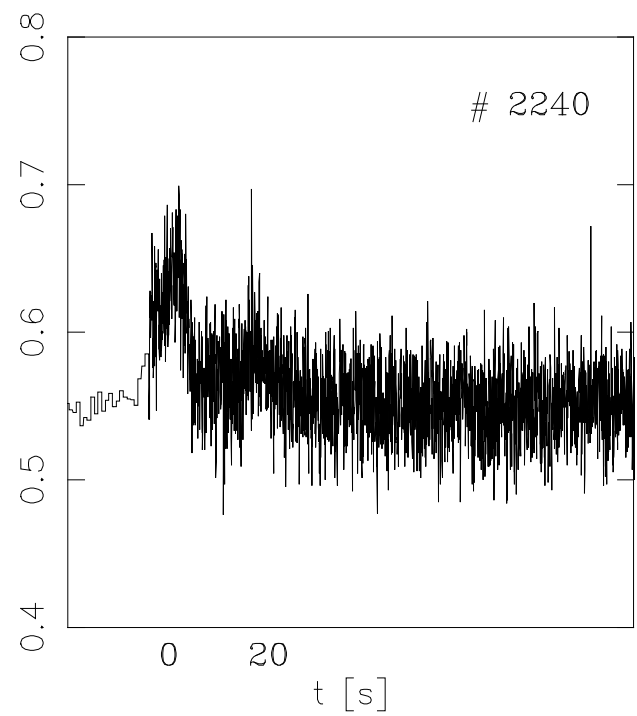
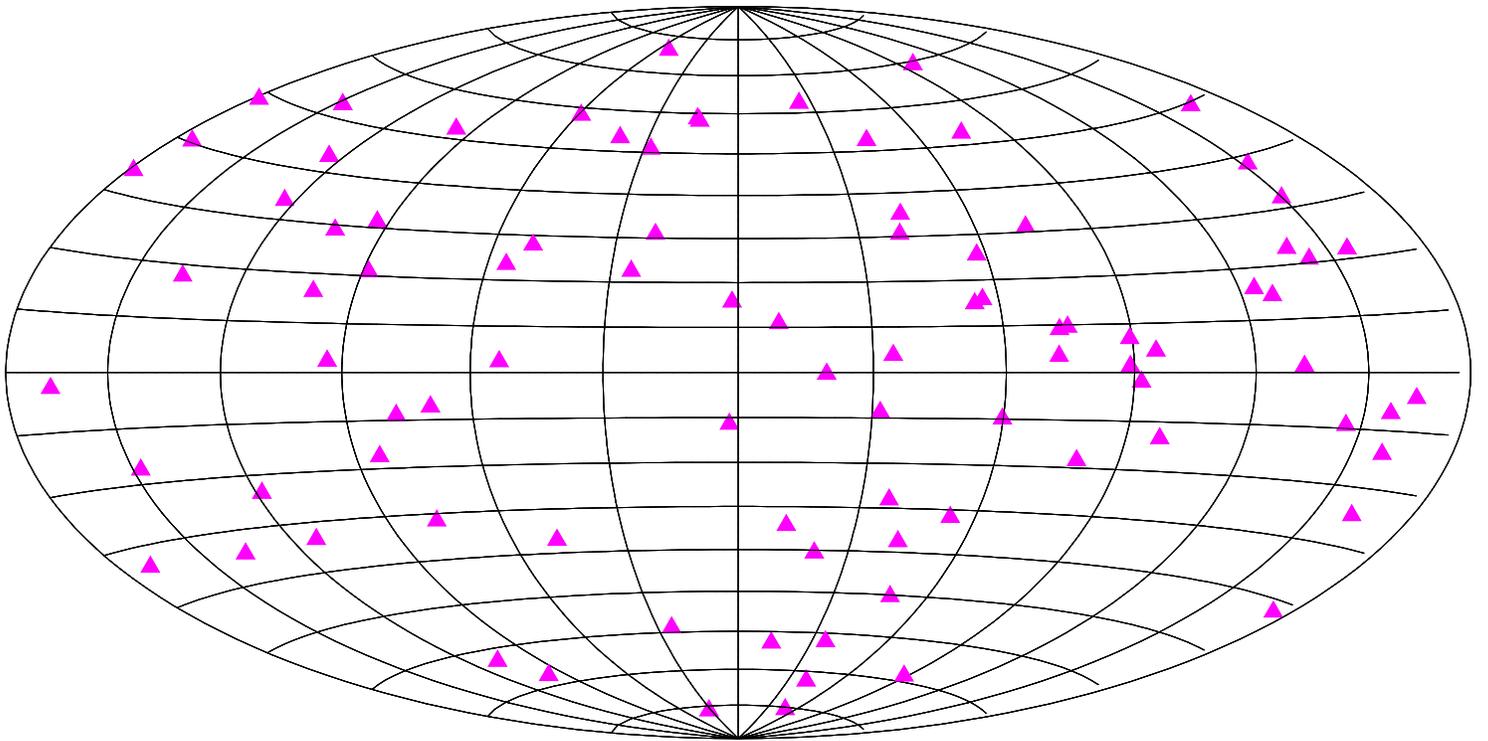
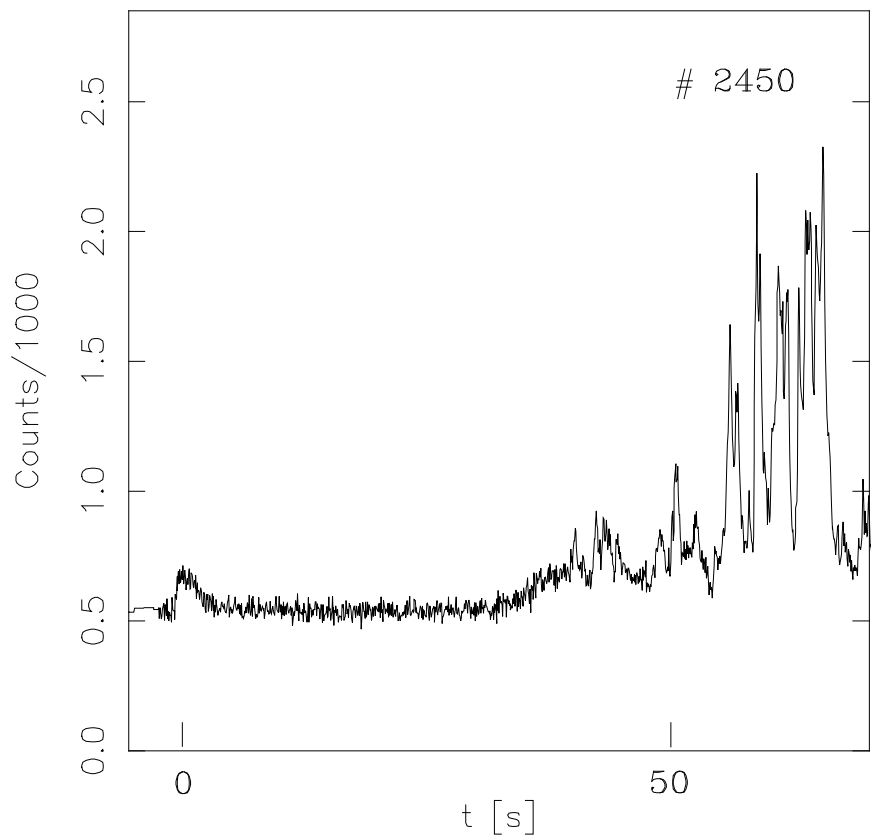
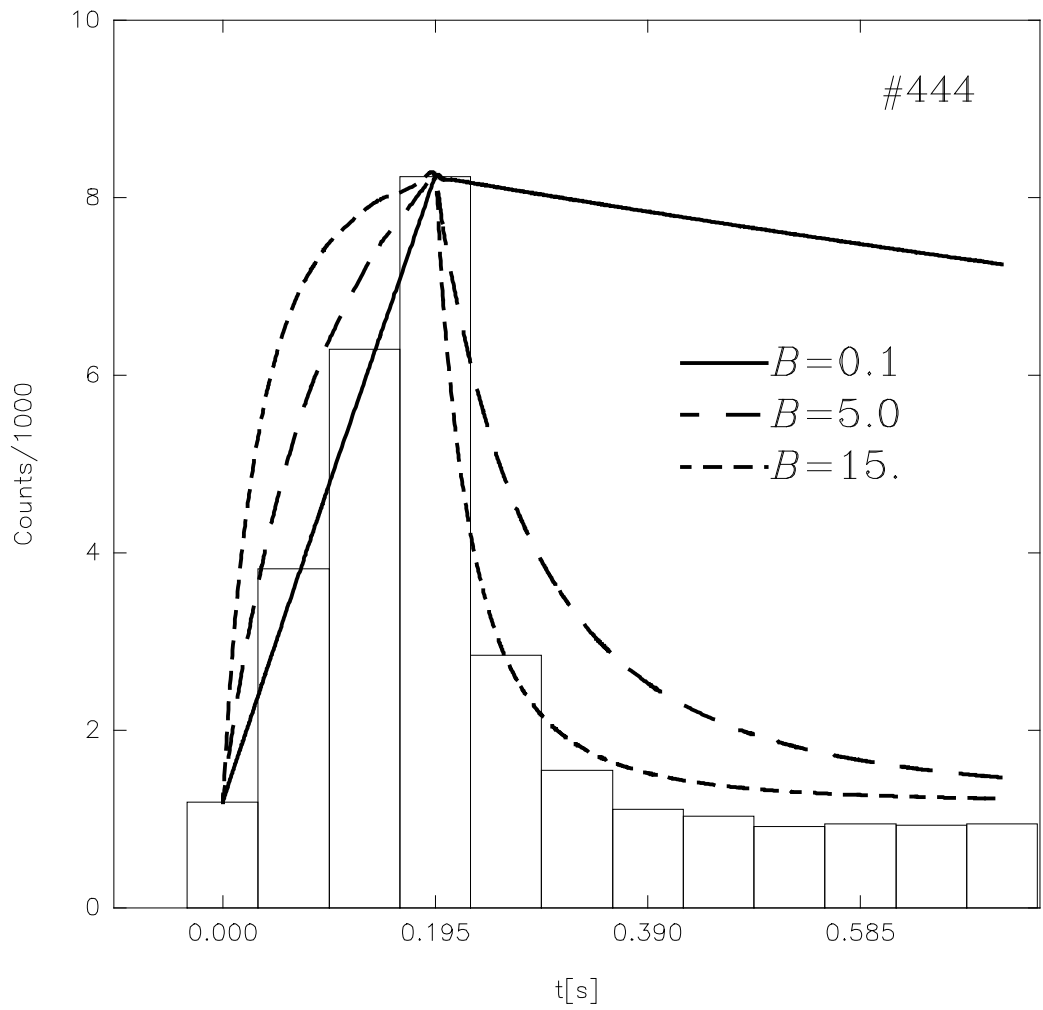
\end{document}